\documentclass[aps,prl,twocolumn,superscriptaddress,showpacs,longbibliography]{revtex4-1}
\usepackage[colorlinks=true,citecolor=blue,linkcolor=blue,breaklinks=true]{hyperref}

\usepackage{amssymb}
\usepackage{graphicx}
\usepackage{amsmath}

\usepackage{times}
\usepackage{color}
\usepackage{subfigure}
\usepackage{setspace}
\usepackage{bm}
\usepackage{ulem}

\begin{document}

\newcommand {\beq} {\begin{equation}}
\newcommand {\eeq} {\end{equation}}
\newcommand {\bqa} {\begin{eqnarray}}
\newcommand {\eqa} {\end{eqnarray}}
\newcommand {\ca} {\ensuremath{c^\dagger}}
\newcommand {\ba} {\ensuremath{b^\dagger}}
\newcommand {\Ma} {\ensuremath{M^\dagger}}
\newcommand {\psia} {\ensuremath{\psi^\dagger}}
\newcommand {\fbar} {\ensuremath{\bar{f}}}
\newcommand {\psita} {\ensuremath{\tilde{\psi}^\dagger}}
\newcommand{\lp} {\ensuremath{{\lambda '}}}
\newcommand{\A} {\ensuremath{{\bf A}}}
\newcommand{\Q} {\ensuremath{{\bf Q}}}
\newcommand{\kk} {\ensuremath{{\bf k}}}
\newcommand{\qq} {\ensuremath{{\bf q}}}
\newcommand{\kp} {\ensuremath{{\bf k'}}}
\newcommand{\rr} {\ensuremath{{\bf r}}}
\newcommand{\rp} {\ensuremath{{\bf r'}}}
\newcommand {\ep} {\ensuremath{\epsilon}}
\newcommand{\lm} {\ensuremath{\lambda^{(1)}}}
\newcommand{\lmm} {\ensuremath{\lambda^{(2)}}}
\newcommand{\rs} {\ensuremath{\rho^{(1)}}}
\newcommand{\rss} {\ensuremath{\rho^{(2)}}}
\newcommand{\nbr} {\ensuremath{\langle r r' \rangle}}
\newcommand {\no} {\nonumber}
\newcommand{\up} {\ensuremath{\uparrow}}
\newcommand{\dn} {\ensuremath{\downarrow}}
\newcommand{\rcol} {\textcolor{red}}
 \begin{abstract}
   Some interacting disordered many-body systems are unable to thermalize when the quenched disorder becomes larger than a threshold value. Although several properties of nonzero energy density eigenstates (in the middle of the many-body spectrum) exhibit a qualitative change across this many-body localization (MBL) transition, many of the commonly-used diagnostics only do so over a broad transition regime. Here, we provide evidence  that the transition can be located precisely even at modest system sizes by sharply-defined changes in the distribution of {\it extremal eigenvalues} of the reduced density matrix of subsystems. In particular, our results suggest that  $p* = \lim_{\lambda_2 \rightarrow \ln(2)^{+}}P_2(\lambda_2)$, where $P_2(\lambda_2)$ is the probability distribution of the second lowest entanglement eigenvalue $\lambda_2$, behaves as an ``order-parameter'' for the MBL phase: $p*> 0$ in the MBL phase, while $p* = 0$ in the ergodic phase with thermalization. Thus, in the MBL phase, there is a nonzero probability that a subsystem is entangled with the rest of the system only via the entanglement of one subsystem qubit with degrees of freedom outside the region. In contrast, this probability vanishes in the thermal phase.
 \end{abstract}

\title{ Extremal statistics of entanglement eigenvalues can track the many-body localized to ergodic transition}
\author{Abhisek Samanta}\email{abhisek@theory.tifr.res.in}
\author{Kedar Damle} \email{kedar@theory.tifr.res.in}
\author{Rajdeep Sensarma}\email{sensarma@theory.tifr.res.in}
 \affiliation{Department of Theoretical Physics, Tata Institute of Fundamental
 Research, Mumbai 400005, India.}

\date{\today}

\maketitle
%

\noindent{\textbf {\textit{Introduction:}}}
Strongly disordered interacting many body systems in one dimension have been predicted to exhibit the absence of transport at finite temperatures, a phenomenon known as many-body localization (MBL)~\cite{basko,mirlin}. Contrary to thermal or ergodic systems, where basic tenets of equilibrium statistical mechanics hold, these systems cannot act as their own heat bath when initialized to arbitrary initial states out of thermal equilibrium~\cite{oganesyan,apal}. Several candidate experimental systems~\cite{mblexpt1,mblexpt2} and theoretical models~\cite{xxz,apal,aubryandre} have been argued to exhibit such a transition from ergodic behaviour to the MBL phase upon increasing the disorder strength~\cite{Altman_review,Huse_review}.

MBL and ergodic phases can be distinguished by the properties of many-body eigenstates in the middle of the spectrum, i.e. with nonzero energy density relative to the ground state~\cite{Altman_review,Huse_review}. In the MBL phase, this eigenspectrum is characterized by energy gaps with a Poisson distribution, violation of the eigenvalue thermalization hypothesis (ETH)~\cite{eth1,eth2}, and short range (area law) entanglement entropy of subsystems even in eigenstates with nonzero energy density. In contrast, ergodic systems show Wigner-Dyson gap statistics, follow the eigenvalue thermalization hypothesis and have long range (volume law) entanglement entropy in the middle of the spectrum~\cite{ee1,ee2}. MBL phases are also characterized by the bimodal nature of the density of entanglement eigenvalues~\cite{ent_spectrum}, and by the logarithmic growth of entanglement entropy~\cite{ee_log} at intermediate time scales upon evolving from an initial Fock state. While the MBL and ergodic phases are clearly distinguished by these contrasting properties, they do not provide a sharp distinction that can be used to precisely locate the transition between the two phases at accessible system sizes. The level statistics, for example,  changes behaviour over a broad transition regime, and has strong finite size effects~\cite{apal}.
\begin{figure}[h!]
 \centering
 \includegraphics[width=0.47\textwidth]{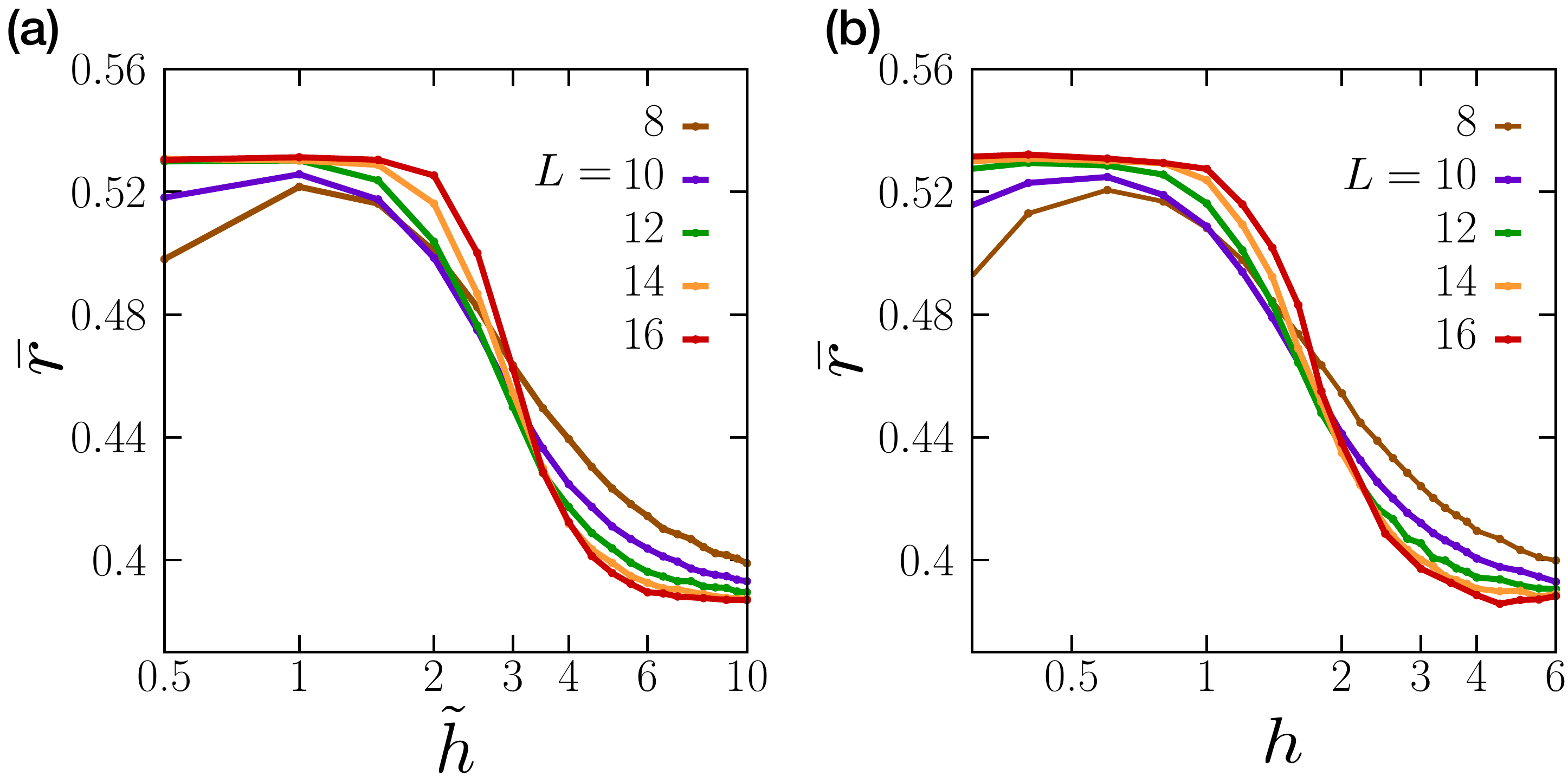}
 \caption{ The average ratio of adjacent many body gaps $\bar{r}$ for the random field Heisenberg model as a function of the disorder strength in the magnetic field for (a) a uniform distribution of fields with width $\tilde h$ and (b) a Gaussian distribution of standard deviation $h$. $\bar{r}$ goes from a Wigner-Dyson value of $0.53$ in the weak disorder limit to a Poissonian value of $0.39$ in the strong disorder limit. Note that the uniform distribution has a critical $\tilde h$ about about twice the critical $h$ for the Gaussian distribution. For each disorder strength, we have used the number of samples ranging from $2\times 10^4$ for $L=8$ to $400$ for $L=16$. }
 \label{rbar_transition}
\end{figure}

The MBL-ergodic  transition has been interpreted earlier~\cite{basko,mirlin} as an Anderson localization transition in Fock space: on the MBL side, the many-body eigenfunctions have support over a few basis states. Heuristically, this can be thought of as a fragmentation of Fock space into small ``clusters'' with weak connectivity between them. In contrast, in the thermal phase, many-body eigenfunctions have support over a more extensive set of Fock states. In this Letter, we explore how this distinction is encoded in the largest few eigenvalues of the reduced density matrix of a subsystem. We focus on the {\it smallest} few entangletment eigenvalues $\lambda_\alpha \equiv -\ln(\rho_\alpha)$, where $\rho_\alpha$ are the {\it largest} few eigenvalues of the subsystem's reduced density matrix corresponding to an eigenstate in the middle of the spectrum. Our key result is that the distribution of these extremal entanglement eigenvalues provides us new information about the nature of the MBL phase, as well as a way of locating the MBL-ergodic transition precisely, with weak finite-size corrections.

Specifically, we propose that  $p* = \lim_{\lambda_2 \rightarrow \ln(2)^{+}}P_2(\lambda_2)$, where $P_2(\lambda_2)$ is the probability distribution of the second lowest entanglement eigenvalue $\lambda_2$, behaves as an ``order-parameter'' for the MBL phase: $p*> 0$ in the MBL phase, while $p* = 0$ in the ergodic phase with thermalization. Thus, in the MBL phase, there is a nonzero probability that a subsystem is entangled with the rest of the system only via the entanglement of one subsystem qubit with degrees of freedom outside the region, while this probability is essentially zero even in the ergodic phase. Indeed, we find that $P_2 \rightarrow 0$ exponentially quickly in the ergodic phase as $\lambda_2$ approaches its lowest possible value of $\ln(2)$. Importantly, this discriminator is already sharp at relatively modest system sizes, and pinpoints the transition point within the somewhat broader transition regime identified by more well-known diagnostics of the transition.

Additionally, we study the functional form of $P_1(\lambda_1)$, the probability distribution of the smallest entanglement eigenvalue $\lambda_1$. 
Although this is, by its very definition, identical 
in the range $\lambda_1 \in [0, \ln(2))$ to the density of entanglement
eigenvalues studied earlier~\cite{ent_spectrum} as a probe of many-body localization, we show that in this range $P_1$ has a characteristic power law, whose exponent can also be used to track the MBL-ergodic transition. We note that this is different from the power law dependence of $\langle\lambda_\alpha\rangle$ with $\alpha$ seen in Ref~\cite{serbyn}.
Finally, we also study the distributions $P_3$ and $P_4$ of the next two lowest entanglement eigenvalues $\lambda_3$ and $\lambda_4$, finding that they too bear a discernible signature of the same MBL to ergodic transition, although lacking precision and clarity of the order parameter $p*$ obtained from $P_2(\lambda_2)$.

\medskip
\noindent{\textbf {\textit{ Model and methods:}}}
We work with spin $S=1/2$ degrees of freedom on a one dimensional lattice of $L$ sites. Our Hamiltonian is the Heisenberg model with random fields, which is the canonical model for studying many body localization,\begin{equation}
 H=\sum_{i=1}^{L}J\vec{S}_i.\vec{S}_{i+1} + h_i S_i^z .
\end{equation}
Here $\vec{S}_i=\frac{1}{2}\vec{\sigma}_i$ and $\vec{\sigma}_i$ are Pauli matrices at each site $i$. The Heisenberg coupling $J$ is set to $1$ throughout this paper. $h_i$ is a random field, drawn independently for each site $i$ from a Gaussian distribution of zero mean and standard deviation $h$, which sets the strength of the disorder; i.e. $P(h_i)=(1/\sqrt{2\pi} h)e^{-h_i^2/2h^2}$. The $z$ component of the total spin $S^z_{tot}$ is a conserved quantity in this model. We numerically diagonalize the Hamiltonian  in the $S^z_{tot}=0$ sector to obtain the eigenvalues $E_n$ and the eigenstates $|\psi_n\rangle$ of the Hamiltonian.

This random field Heisenberg model has been studied before with a uniform distribution of the fields; i.e. $P(h_i)=\Theta(\tilde{h}-h_i)\Theta(h_i)/\tilde{h}$, where it undergoes an ergodic to MBL transition at $\tilde h \approx 3.8$. To make a connection with our model, we define the gap between successive eigenstates $\delta_n =E_{n+1}-E_n$. The ratio of successive gaps, $r_n ={\mathrm{Min}}(\delta_n,\delta_{n+1})/{\mathrm{Max}}(\delta_n,\delta_{n+1})$ is then averaged over eigenstates and disorder realizations to obtain $\bar{r}$, which interpolates between its GOE value of $0.53$ in the ergodic phase, to its Poissonian value of $0.39$ in the MBL phase. In Fig.~\ref{rbar_transition}(a) and (b) we plot $\bar{r}$ as a function of $\tilde h$ and $h$ for the uniform and the Gaussian distribution respectively. From the crossing of the curves at largest system sizes, the transition in the uniform distribution occurs at $\tilde h \approx 3.5$ in the uniform case and at $h \approx 1.8$ for the Gaussian distribution, although it shows significant variations with the system size. The factor of $2$ between the uniform and Gaussian distribution is explained by the fact that the Gaussian distribution gives a finite probability for very large values of $h_i$, whereas the uniform distribution cuts off the possible values of $h_i$. From Fig.~\ref{rbar_transition}(b), one can clearly say that $h=0.5$ is deep in the ergodic phase and $h=6.0$ is deep in the MBL phase. These are the canonical values we will use to describe the two phases. The $\bar{r}$ curve shows a region of $h=1$ to $h=3$ as the transition region for the Gaussian distribution.

We consider a subsystem of size $L_A$ and construct the density matrix $\hat{\rho}^{(n)}$  for this subsystem from each of the eigenstates $|\psi_n\rangle$ by tracing out degrees of freedom in the rest of the system. The eigenvalues of the density matrix $\rho^{(n)}_\alpha$ are related to the entanglement eigenvalues $\lambda^{(n)}_\alpha=-\ln \rho^{(n)}_\alpha$, where $\alpha = 1,2.. 2^{L_A}$, and the eigenvalues $\rho^{(n)}_\alpha$ are arranged in descending order of values, i.e. $\rho^{(n)}_1$ is the largest eigenvalue and so on, with  the sum-rule $\sum_\alpha \rho^{(n)}_\alpha=1$. Consequently the entanglement eigenvalues are arranged in ascending order, i.e. $\lambda^{(n)}_1$ is the lowest entanglement eigenvalue and so on. We tabulate the lowest four entanglement eigenvalues obtained from each of the eigenstates (in the middle one third of the spectrum) for different disorder realizations to construct the distributions $P_\alpha(\lambda_\alpha)$ for $\alpha=1,2,3,4$. As noted earlier, $P_1$ in the range $\lambda_1 \in [0,\ln(2))$ is identical to the density of entanglement eigenvalues, but contains new information for $\lambda_1 > \ln(2)$.

  Indeed, the overall density of entanglement eigenvalues constructed by averaging over states in the middle of the spectrum is affected by the trace constraint on $\rho^{(n)}_\alpha$ in a manner that is hard to disentangle: This density has contributions from all the entanglement eigenvalues obtained from a single state $n$, in addition to contributions from other states. Contributions from a single state are constrained by a sum rule since $\sum_i\rho^{(n)}_\alpha = 1$, whereas $\rho^{(n)}_\alpha$ for different $n$ are not similarly constrained by each other's values. This is one of our motivations for focusing on the distributions of the smallest few entanglement eigenvalues, which are therefore expected to have sharper signatures of the underlying localization phenomenon.
\begin{figure}[h!]
	\centering
	\includegraphics[width=0.485\textwidth]{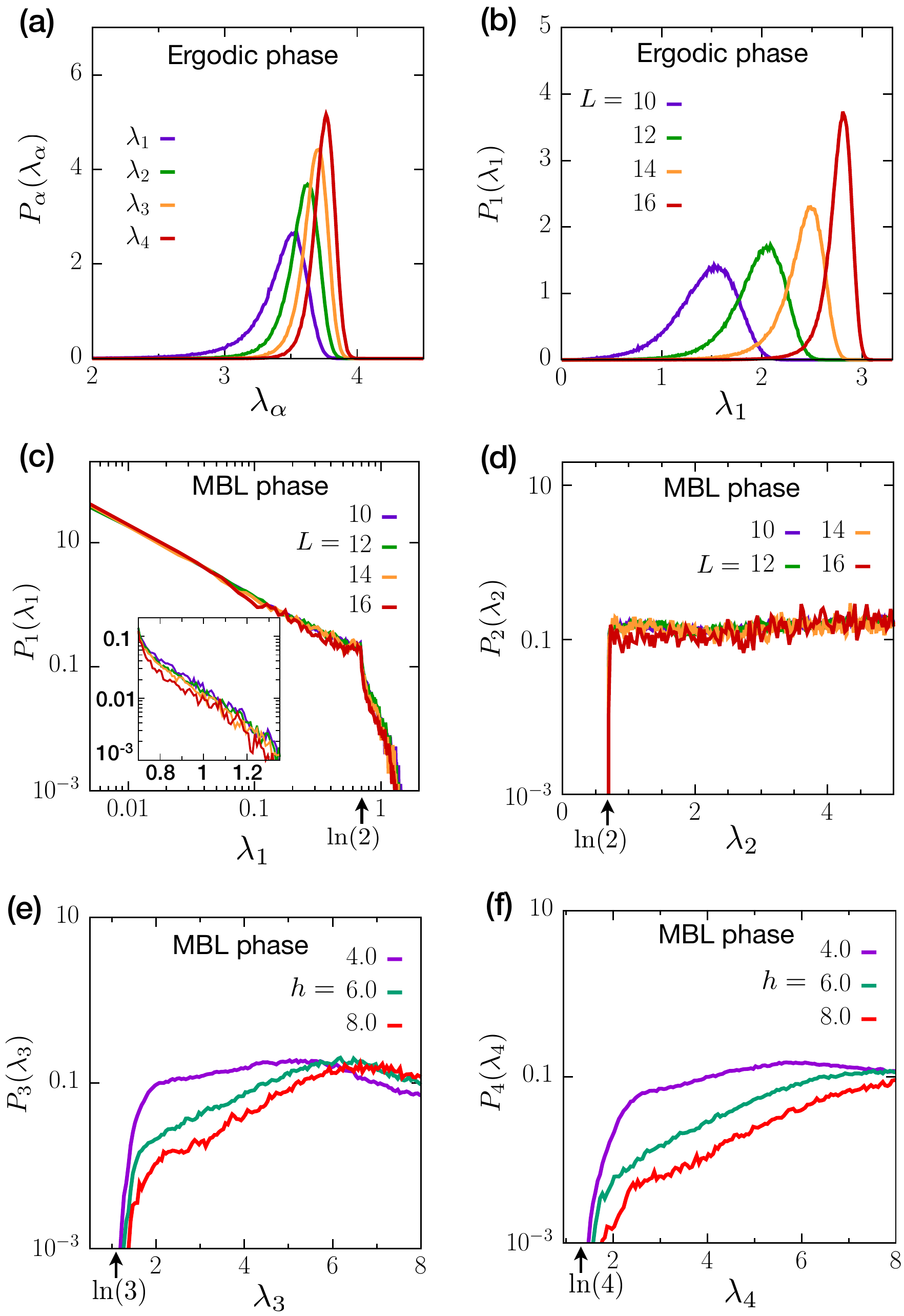}
	\caption{ (a): Distribution of lowest four entanglement eigenvalues $\lambda_1, ..\lambda_4$ in the ergodic phase ($h=0.5$) for system size $L=16$ and subsystem size $L_A=8$. They all have similar skew symmetric structures, while the distribution becomes sharper and the peak position increases from $\lambda_1$ to $\lambda_4$. (b)-(c): Distribution of lowest entanglement eigenvalue $\lambda_1$ for fixed subsystem size $L_A=5$ and system size $L=10, 12, 14$ and 16. (b) Distribution in ergodic phase ($h=0.5$) showing the peak position increases and the distribution sharpens with increasing $L$ for fixed $L_A$. (c) Distribution in MBL phase ($h=6.0$) showing power law upto $\ln(2)$. Inset shows exponential distribution beyond $\ln(2)$. The distribution is independent of $L$. (d): Distribution of second lowest entanglement eigenvalue $\lambda_2$ in the MBL phase ($h=6.0$) showing a finite value of the distribution at $\ln(2)$ for $L_A=5$. The distribution is independent of $L$. (e)-(f): Distribution of (e) third lowest and (f) fourth lowest entanglement eigenvalues for $L=16$ and $L_A=8$ in the MBL phase ($h=4.0, 6.0$ and 8.0). Note that the distributions go to zero at $\ln(3)$ and $\ln(4)$ respectively.}
	\label{plambda_phase}
\end{figure}

\medskip
\noindent{\textbf {\textit{The Ergodic phase:}}}
We first consider the distribution of low-lying entanglement eigenvalues at a weak disorder of $h=0.5$, where, as seen from Fig.~\ref{rbar_transition}(b), the system is deep in the ergodic phase. For the delocalized eigenstates with volume law entanglement entropy, one would expect the typical low lying entanglement eigenvalues to strongly depend on the system and subsystem size. In Fig.~\ref{plambda_phase}(a), we plot the distribution of the lowest four entanglement eigenvalues $\lambda_1 .. \lambda_4$ for system size $L=16$ and subsystem size $L_A=8$. The distribution $P_\alpha(\lambda_\alpha)$ are characterized by a sharp peak at $\lambda^p_\alpha$ with a skew symmetric tail, which is broader on the left than on the right. Empirically, the large deviation function on the right (i.e. $\lambda_\alpha>\lambda^p_\alpha$) is given by $P_\alpha(\lambda_\alpha) \sim e^{-\left[\frac{\lambda_\alpha-\lambda^p_\alpha}{l_R}\right]^3}$, where as the tail on the left  (i.e. $\lambda_\alpha<\lambda^p_\alpha$) is given by $P_\alpha(\lambda_\alpha) \sim e^{-\left[\frac{|\lambda_\alpha-\lambda^p_\alpha|}{l_L}\right]^{3/2}}$. Fig.~\ref{plambda_phase}(b) shows the distribution of the lowest entanglement eigenvalue $P_\alpha(\lambda_1)$ for a fixed $L_A=5$ for different system sizes $L=10,12,14,16$. The peak position scales linearly with the size of the system, while the distribution narrows with increasing system size in the ergodic phase of the system. Similar finite size effects are also seen in the distribution of other low lying entanglement eigenvalues~\cite{samanta}.

\medskip
\noindent{\textbf {\textit{The MBL phase:}}} We now consider the distributions of low-lying entanglement eigenvalues at a strong disorder of $h=6.0$, where the system is deep in the many body localized phase. The distributions show several distinctive features in this case.

We first focus on the distribution $P_1$ of the lowest entanglement eigenvalue $\lambda_1$. The distribution function $P_1(\lambda_1)$ is plotted in Fig.~\ref{plambda_phase}(c) on a log-log plot for a fixed subsystem size $L_A=5$ and different system sizes $L=10,12,14,16$. As expected, deep in the localized phase, the distribution function is insensitive to system size. There is a large weight in the limit $\lambda_1 \rightarrow 0$, corresponding to the occurrence of product states in the MBL phase. The distribution has a power-law form for small $\lambda_1 \ll \ln (2)$. Close to $\lambda_1=0$, this power-law divergence is cutoff in a characteristic manner~\cite{samanta}. Beyond $\lambda_1=\ln(2)$,  the distribution decays exponentially, as seen in the semi-log plot  in the inset of Fig.~\ref{plambda_phase}(c). This leads to a distinctive kink in the distribution function at $\lambda_1=\ln(2)$; i.e.
\begin{eqnarray}
\no P_1[\lambda_1]&\sim& [\lambda_1]^{-b} ~~~ \text{for} ~~ 0 < \lambda_1 <\ln(2)\\
& \sim&e^{-\lambda_1/\lambda_0} ~~~~~~~~~~~~~~~ ~~ \lambda_1 > \ln (2)
\end{eqnarray}
or equivalently for the eigenvalues of the density matrix,
\begin{eqnarray}
\no P_1[\rho_1]&\sim& \frac{1}{\rho_1 [-\ln(\rho_1)]^b}  ~~~ \text{for} ~~ \frac{1}{2}\leq \rho_1 <1\\
& \sim&[\rho_1]^{\frac{1}{\lambda_0}-1} ~~~~~~~~~~~~~~~ ~~ \frac{1}{2}>\rho_1 > 0
\end{eqnarray}

Naturally, the power law behaviour we find for $\lambda_1 \rightarrow 0$ matches the known behaviour~\cite{ent_spectrum} of the density of entanglement eigenvalues in this range. As noted earlier, it can be understood in terms of presence of local integrals of motion (LIOM), whose weight decay exponentially with distance from its central location~\cite{serbyn}. 

We now turn our attention to the distribution of $\lambda_2$. By definition, $\lambda_2 \in (\ln(2), L_A \ln(2))$. The distribution $P_2(\lambda_2)$ for $h=6.0$ is plotted in Fig.~\ref{plambda_phase}(d) for a fixed $L_A=5$ for $L=10,12,14,16$. 
At large $\lambda_2$ (not shown in figure), it decays exponentially. $\lambda_2= \ln(2)$ is a special value which corresponds to $\rho_1=\rho_2=1/2$, and $\rho_\alpha=0$ for $\alpha>2$. The many body eigenstate which gives rise to this can be decomposed as $|\psi\rangle = \frac{1}{\sqrt{2}}[ |\psi^1_A\rangle |\psi^1_B\rangle+|\psi^2_A\rangle |\psi^2_B\rangle]$, where $|\psi^i_{A(B)}\rangle$ are respectively states in the Hilbert space of subsystems $A$ and the rest of the system $B$.  In other words, exactly one subsystem qubit is maximally entangled with one degree of freedom from the environment. 
\begin{figure}[h!]
	\centering
	\includegraphics[width=0.485\textwidth]{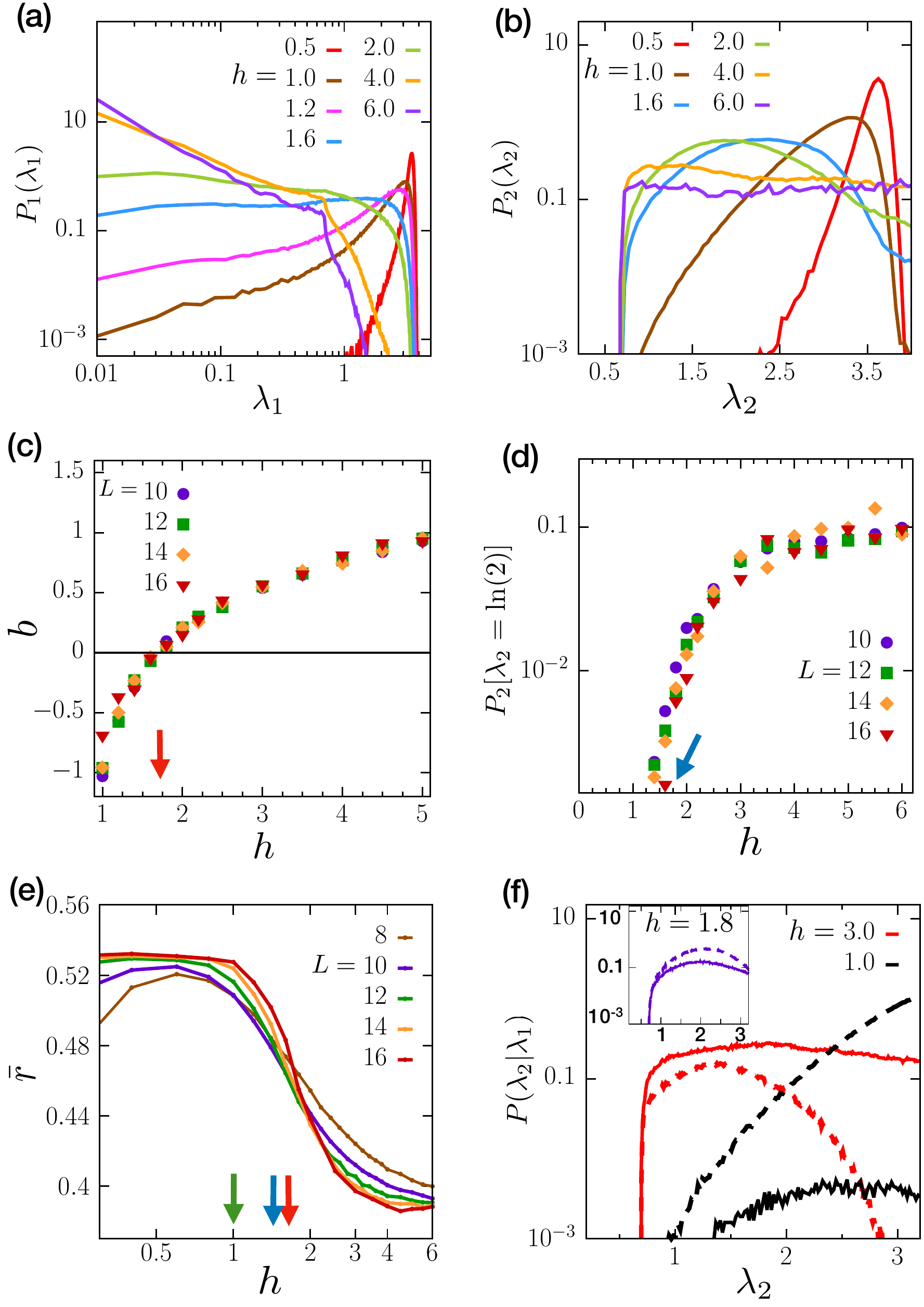}
	\caption{ (a)-(b): Change in the distribution of (a) lowest entanglement eigenvalue $\lambda_1$ and (b) second lowest entanglement eigenvalue $\lambda_2$ with $h$ from ergodic to MBL phases for system size $L=16$ and subsystem size $L_A=8$. (c)-(d): Features of extremal entanglement eigenvalue distribution are used to track the MBL-ergodic transition for fixed subsystem size $L_A=5$ and system size $L=10, 12, 14$ and 16. (c) The exponent $b$ of the power law distribution, $P_1(\lambda_1) \sim (\lambda_1) ^{-b}$ as a function of $h$. The value of $h$ where $b$ changes sign is taken as $h_c=1.75$ (indicated by red arrow). The answers are very weakly dependent on system size $L$. (d) The probability of getting a maximally entangled qubit, $P_2[\lambda_2=\ln(2)]$ as a function of $h$. The probability goes to zero at $h_c \sim 1.3-1.6$ (indicated by blue arrow). The answers are weakly dependent on system size $L$. (e): The location of $h_c$ obtained from above definitions with the gap statistics curve. The definitions using the exponent $b$ and $P_2[\lambda_2=\ln(2)]$ lie in the middle of the transition region. The definition using $P_1[\lambda_1=0]$ does not track the transition (indicated by the green arrow). (f): The conditional distribution of $\lambda_2$ conditioned on $\lambda_1 < \ln(2) $ (solid lines) and  $\lambda_1 > \ln(2) $ (dashed lines) for $h=3.0, 1.0$. The conditional distributions change their relative magnitude between these values; Inset: the conditional distributions reach same limiting value at $\ln(2)$ at $h=1.8$. In Fig.~\ref{plambda_phase} and \ref{plambda_transition}, we have used the number of samples ranging from $3600$ for $L=10$ to $400$ for $L=16$.}
	\label{plambda_transition}
\end{figure}

We finally consider the distribution of the $3^{rd}$ and $4^{th}$ lowest entanglement eigenvalues. We note that $\ln (3) < \lambda_3 < L_A \ln (2)$ and $\ln (4) < \lambda_4 < L_A \ln (2)$ by their definitions. The distributions of $\lambda_3$ and $\lambda_4$ are plotted in Fig.~\ref{plambda_phase}(e) and (f) respectively for $h=4.0,6.0,8.0$. It is clear from the figure that $P_3[\lambda_3=\ln(3)]$ and $P_4[\lambda_4=\ln(4)]$ tend to zero: The distributions decrease with decreasing $\lambda_\alpha$, before a final sharp drop on the left edge. We note that even if one ignores the final drop and extrapolates the curve, these extrapolated values of $P_3(\ln(3))$ and $P_4(\ln(4))$ keep decreasing with increasing $h$, so that deep in the MBL phase these probabilities extrapolate to zero. Note that $\lambda_\alpha=\ln(\alpha)$ corresponds to a maximally entangled state in a space spanned by $\alpha$ basis states. Therefore, we conclude that such an entanglement structure for $\alpha=3$ and $\alpha=4$ arises with exponentially small probability in the MBL phase. This is consistent at a heuristic level with the area-law expected for the entanglement entropy, which essentially measures the number of such pairs of qubits entangled across the boundary of the system.

The distribution of the lowest four entanglement eigenvalues thus paints the following picture of the entanglement structure in the MBL phase: With nonzero probability, a subsystem is entangled with the surroundings only due to the maximal entanglement of one subsystem qubit with one degree of freedom from the surrounding, whereas higher order entanglement structures, in which the subsystem exchanges multiple bits of information, are exponentially suppressed.

\medskip
\noindent{\textbf {\textit{The MBL-Ergodic transition:}}} We now focus our attention on the transition between the MBL and ergodic phases.  Let us first consider the distribution of $\lambda_2$ (for $L=16$ and $L_A=8$), which is plotted for different values of $h$ across the transition in Fig.~\ref{plambda_transition}(b). Clearly, $P_2[\lambda_2= \ln(2)]$ saturates at large $h$ and vanishes below a critical disorder. 
The vanishing of $P_2[\lambda_2= \ln(2)]$ can therefore serve as a precise indicator of the MBL-ergodic transition; i.e. $P_2[\lambda_2= \ln(2)]$ can act as an ``order-parameter'' for the transition. In other words, the MBL phase is characterized by a finite probability of exactly $1$ bit of information exchanged between the subsystem and the surrounding. In Fig.~\ref{plambda_transition}(d) we plot $P_2[\lambda_2= \ln(2)]$  as a function of $h$ for different system size, which gives a critical disorder $h\sim 1.3-1.6$. This value is shown in Fig.~\ref{plambda_transition}(e) by a blue arrow. We note that the vanishing of  $P_2[\lambda_2= \ln(2)]$ gives a transition point in the center of the transition region defined by variation of $r$. The shift of the transition point with system size is much smaller with our new criterion than with more traditional indicators of the transition. This is our key observation.

Next, we consider the distribution of $\lambda_1$. For $L=16$ and $L_A=8$, this  is plotted for different values of  $h$ across the transition in Fig.~\ref{plambda_transition}(a). As the system goes through the MBL-ergodic transition, the distribution at small $\lambda_1$ can still be fitted to a power law~\cite{samanta}, but the exponent $b$ changes from positive at large disorder to negative at small disorder. The kink at $\ln(2)$ also vanishes as $b$ changes sign. The extracted value of the exponent $b$ (using $L_A=5$) is plotted as a function of $h$ for different values of system size in Fig.~\ref{plambda_transition}(c). We note that the finite size variation in $b$ is small.  One criterion that can be used to track the MBL-ergodic transition is the sign change in $b$. From Fig.~\ref{plambda_transition}(c), we get $h=1.75$ as the critical disorder for the transition, where the distribution in Fig.~\ref{plambda_transition}(a) shows a horizontal part.


We note that a similar criterion of sign change of exponent of the power law in $\langle\lambda_\alpha\rangle$ vs $\alpha$ was proposed in Ref.~\cite{serbyn}. However, the finite size variation of $h_c$ is much smaller if one uses the power law exponent of $P_1(\lambda_1)$.
The  critical value we obtain is indicated as a red arrow in Fig.~\ref{plambda_transition}(e), where the gap statistic $r$ is plotted as a function of $h$. The critical value lies in the center of the transition defined by the variation of $r$.
The vanishing of $P_1[\lambda_1=0]$ can also be taken as an alternate definition for the transition, corresponding to the absence of product states among the eigenstates. However, from Fig.~\ref{plambda_transition}(a), this occurs at a much lower value of $h$ than the sign change of $b$. This value is indicated in Fig.~\ref{plambda_transition}(e) by a green arrow, and corresponds to the deviation from Poissonian statistics rather than the center of the transition region.

To gain further insight into the distribution of entanglement eigenvalues across the transition, we consider the conditional probability distribution $P[\lambda_2| \lambda_1 <\ln(2)]$ (shown in Fig.~\ref{plambda_transition}(f) by solid lines) and $P[\lambda_2| \lambda_1 >\ln(2)]$ (shown in Fig.~\ref{plambda_transition}(f) by dashed lines). In the MBL phase ($h=3.0$), the solid line goes over the dashed line, since $P_1(\lambda_1)$ is decreasing around $\ln (2)$. In the ergodic phase $P_1(\lambda_1)$ is increasing around $\ln (2)$, and hence the dashed line overshoots the solid line. At $h=1.8$, close to the transition, the two conditional probabilities come very close to each other.

\medskip
\noindent{\textbf {\textit{ Discussion:}}}
In summary, in this paper, we have shown a new way to characterize the MBL and ergodic phases, which can be used to pinpoint the MBL-ergodic transition. In particular, we have studied the distribution of the lowest two entanglement eigenvalues. The sign change of the power law exponent of the distribution of lowest entanglement eigenvalue tracks the critical disorder. This can also be obtained by considering the the disorder where the probability of the second lowest entanglement eigenvalue being $\ln(2)$ becomes finite. Thus  the probability that a subsystem exchanges exactly one bit of information with the surroundings can be used as an ``order parameter'' for the MBL-ergodic transition.

{\bf Acknowledgements:}
The authors acknowledge the computational facilities at the Department of Theoretical Physics, TIFR Mumbai for this work.

\clearpage

\begin{thebibliography}{99} 

\bibitem{basko} D. M. Basko, I. L. Aleiner, and B. L. Altshuler, Metal-insulator transition in a weakly interacting many-electron system with localized single-particle states, \textit{Ann. Phys.} {\bf 321}, 1126 (2006).

\bibitem{mirlin} I. V. Gornyi, A. D. Mirlin, and D. G. Polyakov, Interacting electrons in disordered wires: Anderson localization and low-$T$ transport, \textit{Phys. Rev. Lett.} {\bf 95}, 206603 (2005).

\bibitem{oganesyan} V. Oganesyan, and D. A. Huse, Localization of interacting fermions at high temperature, \textit{Phys. Rev. B} {\bf 75}, 155111 (2007).

\bibitem{apal} A. Pal, and D. A. Huse, Many-body localization phase transition, \textit{Phys. Rev. B} {\bf 82}, 174411 (2010).

\bibitem{mblexpt1} M. Schreiber, S. S. Hodgman, P. Bordia, H. P. Luschen, M. H. Fischer, R. Vosk, E. Altman, U. Schneider, and I. Bloch, Observation of many-body localization of interacting fermions in a quasirandom optical lattice, \textit{Science} {\bf 349}, 842 (2015).

\bibitem{mblexpt2} J. Smith, A. Lee, P. Richerme, B. Neyenhuis, P. W. Hess, P. Hauke, M. Heyl, D. A. Huse, and C. Monroe, Many-body localization in a quantum simulator with programmable random disorder, \textit{Nature Physics} {\bf 12}, 907 (2016).

\bibitem{xxz} M. Znidaric, T. Prosen, and P. Prelovsek, Many-body localization in the Heisenberg $XXZ$ magnet in a random field, \textit{Phys. Rev. B} {\bf 77}, 064426 (2008).

\bibitem{aubryandre} S. Iyer, V. Oganesyan, G. Refael, and D. A. Huse, Many-body localization in a quasiperiodic system, \textit{Phys. Rev. B} {\bf 87}, 134202 (2013).

\bibitem{Altman_review} E. Altman, Many-body localization and quantum thermalization, \textit{Nature Physics} {\bf 14}, 979 (2018).

\bibitem{Huse_review} R. Nandkishore, and D. A. Huse, Many-body localization and thermalization in quantum statistical mechanics, \textit{Annu. Rev. Condens. Matter Phys.} {\bf 6}, 15 (2015).
 
\bibitem{eth1} J. M. Deutsch, Quantum statistical mechanics in a closed system, \textit{Phys. Rev. A} {\bf 43}, 2046 (1991).

\bibitem{eth2} M. Srednicki, Chaos and quantum thermalization, \textit{Phys. Rev. E} {\bf 50}, 888 (1994).


\bibitem{ee1} B. Bauer, and C. Nayak, Area laws in a many-body localized state and its implications for topological order, \textit{J. Stat. Mech.} {\bf 2013}, P09005 (2013).

\bibitem{ee2} J. A. Kjall, J. H. Bardarson, and F. Pollmann, Many-body localization in a disordered quantum Ising chain, \textit{Phys. Rev. Lett.} {\bf 113}, 107204 (2014).

\bibitem{ent_spectrum} S. D. Geraedts, N. Regnault, and R. M. Nandkishore, Characterizing the many-body localization transition using the entanglement spectrum, \textit{New J. Phys.} {\bf 19}, 113021 (2017).

\bibitem{ee_log} J. H. Bardarson, F. Pollmann, and J. E. Moore, Unbounded growth of entanglement in models of many-body localization, \textit{Phys. Rev. Lett.} {\bf 109}, 017202 (2012).

\bibitem{serbyn} M. Serbyn, A. A. Michailidis, D. A. Abanin, and Z. Papic, Power-law entanglement spectrum in many-Body localized phases, \textit{Phys. Rev. Lett.} {\bf 117}, 160601 (2016).

\bibitem{samanta} A. Samanta, K. Damle, and R. Sensarma, Unpublished.

%
%
%
%
%









\end{thebibliography}
\end{document}